\documentclass[12pt]{article}

\usepackage{amssymb}
\usepackage{citesort}
\usepackage{amsmath}
\usepackage{color,graphicx}
\usepackage{epstopdf}

\def\pa{\partial}
\def\fr{\frac}

\def\ii{\textrm i}
\def\ee{\textrm e}

\def\bn{{\boldsymbol \nabla}}

\newcommand{\db}{de$\,$Broglie}

\newcommand{\dbb}{de$\,$Broglie-Bohm}

\newcommand{\be}{\begin{equation}}
\newcommand{\en}{\end{equation}}

\begin{document}

\title{On quantum potential dynamics}

\author{
Sheldon Goldstein\footnote{Departments of Mathematics, Physics and
     Philosophy, Rutgers University, Hill Center,  
     110 Frelinghuysen Road, Piscataway, NJ 08854-8019, USA.
     E-mail: oldstein@math.rutgers.edu} \ and
Ward Struyve\footnote{Departments of Mathematics and Philosophy,
     Rutgers University, Hill Center,  
     110 Frelinghuysen Road, Piscataway, NJ 08854-8019, USA.
     E-mail: wstruyve@math.rutgers.edu}
}
\maketitle

\begin{abstract}
\noindent
Non-relativistic \dbb\ theory describes particles moving under the guidance of the wave function. In \db's original formulation, the particle dynamics is given by a first-order differential equation. In Bohm's reformulation, it is given by Newton's law of motion with an extra potential that depends on the wave function---the quantum potential---together with a constraint on the possible velocities. It was recently argued, mainly by numerical simulations, that relaxing this velocity constraint leads to a physically untenable theory. We provide further evidence for this by showing that for various wave functions the particles tend to escape the wave packet. In particular, we show that for a central classical potential and bound energy eigenstates the particle motion is often unbounded. This work seems particularly relevant for ways of simulating wave function evolution based on Bohm's formulation of the \dbb\ theory. Namely, the simulations may become unstable due to deviations from the velocity constraint.  
\end{abstract}

\renewcommand{\baselinestretch}{1.1}
\bibliographystyle{unsrt}

Non-relativistic \dbb\ theory (also called Bohmian mechanics) \cite{bohm93,holland93b,duerr09} describes point-particles moving under the guidance of the wave function. In the case of spinless particles, with positions ${\bf X}_k$, $k=1,\dots,n$, and configuration $X=({\bf X}_1,\dots, {\bf X}_N)$, the equations of motion are given by 
\begin{equation}
\fr{d {\bf X}_k(t)}{dt}= \frac{1}{m_k} {\boldsymbol {\nabla}}_k S(X(t),t) \,,
\label{1}
\end{equation}
where the wave function $\psi=|\psi(x,t)|\ee^{\ii S(x,t)/\hbar}$, with $x=({\bf x}_1,\dots,{\bf x}_n)$, satisfies the non-relativistic Schr\"odinger equation
\begin{equation}
\ii \hbar \partial_t \psi =  \left( -\sum^n_{k=1} \frac{\hbar^2 }{2m_k}\nabla^2_k + V \right)  \psi \,.
\label{2}
\end{equation}
The theory reproduces the standard quantum predictions provided that for an ensemble of systems all with the same wave function $\psi$ the particle distribution is given by $|\psi|^2$. This distribution---the {\em quantum equilibrium distribution}---is preserved by the particle dynamics, i.e., it does not change its form as a function of $\psi$.

This theory was originally discovered by \db\ in the late 20's \cite{debroglie28} and rediscovered by Bohm in the early 50's \cite{bohm52a,bohm52b}. Unlike \db, Bohm did not regard the equation of motion \eqref{1} as fundamental. Instead, he proposed the second-order differential equation
\begin{equation}
m_k \fr{d^2 {\bf X}_k(t)}{dt^2}= - {\boldsymbol \nabla}_k (V + Q)(X(t),t)  ,
\label{3}
\end{equation} 
which is Newton's law of motion with an extra $\psi$-dependent potential $Q$---the {\em quantum potential}---given by
\begin{equation}
Q = - \sum^N_{k=1} \frac{\hbar^2}{2m_k} \frac{\nabla^2_k |\psi|}{|\psi|} . 
\label{4}
\end{equation} 
The equation \eqref{3} is referred to as Bohm's dynamics in \cite{colin13b}. We prefer to call it the {\em quantum potential dynamics} (QPD for short). In addition, Bohm assumed the constraint 
\be
\frac{ d {\bf X}_k }{dt}(0) = \frac{1}{m_k} \bn_k S(X(0),0)
\label{5}
\en
on the initial velocities. The QPD implies that this constraint is preserved in time, i.e., if \eqref{5} holds then $d {\bf X}_k(t) / dt = \bn_k S(X(t),t)/m_k$ for all times $t$.   

Bohm thought modifications of his theory might be required in order to understand phenomena over distances smaller than $10^{-13}$cm. In particular, he entertained the idea of relaxing the constraint on the velocities \cite{bohm52a}. However, he did not suggest this relaxation without further modifications of the theory. Rather, he considered either modifying the Schr\"odinger equation or the Newtonian equation in order to ensure that over time arbitrary initial velocities would tend to those given by the \dbb\ theory. Thus his modified QPD could reasonably be expected to yield predictions in at least approximate agreement with those of standard quantum mechanics. (And while we are here ignoring spin, it should be noted that the formulation of the QPD for particles with spin is problematical \cite{bohm93}.)

In \cite{colin13b} the possibility was considered of relaxing the constraint on the initial velocities {\em without} additional modifications to the equations of motion. Numerical simulations were performed for the one-dimensional harmonic oscillator and for the non-relativistic hydrogen atom, for particular superpositions of energy eigenstates. In the case of the harmonic oscillator it was found that for initial positions outside the bulk of the wave packet (i.e., where $|\psi|^2$ is appreciably different from zero) and for sufficiently large initial velocities the particles seem to escape to infinity. In addition, an example was provided of a phase-space distribution corresponding to initial momenta larger than those of the \dbb\ theory for which the particles seem to be escaping. It was further argued analytically that for initial positions $x \to \infty$ and for large enough initial velocities the particles escape to infinity. (However, unless the particles can move so far outside the bulk of the packet, in which case the theory seems no good in the first place, it seems that one can ignore such initial conditions.) In the case of the hydrogen atom, a couple of trajectories were simulated and it was found that the more the initial velocity deviates from \eqref{5}, the more the particle seems to escape from the packet. 

Here we provide further evidence, in the form of analytical results, that relaxing the constraint on the velocities is untenable. We consider some potentials $V$ and quantum states for a single particle for which the QPD can be easily analysed. In particular, we consider the case of central potentials and bound energy eigenstates (for which $|\psi|^2$ is appreciably different from zero only in a certain region of space) and show that the particle motion is often unbounded, so that particles escape to infinity. This implies that the QPD is empirically inadequate. Perhaps more importantly, this work also seems to reveal a potential source of instabilities in particular ways of simulating wave function evolution. We will explain this further near the end of the paper. 

Let us first consider some properties of energy eigenstates. For a single particle, the Schr\"odinger equation implies
\begin{equation}
\pa_t S + \frac{1}{2m} {\boldsymbol {\nabla}}  S \cdot {\boldsymbol {\nabla}} S + V + Q = 0 \,.
\label{11}
\end{equation} 
Hence for an energy eigenstate $\psi = \phi({\bf x}) \ee^{-\ii Et/\hbar}$ we have
\begin{equation}
E = \frac{1}{2m} {\boldsymbol {\nabla}} S \cdot {\boldsymbol {\nabla}} S  + V + Q 
\label{12}
\end{equation} 
and the total force in the QPD is given by 
\begin{equation}
- {\boldsymbol \nabla} (V + Q) = - \frac{1}{2m} {\boldsymbol \nabla} \left( {\boldsymbol {\nabla}}  S \cdot {\boldsymbol {\nabla}} S \right) \,.
\label{13}
\end{equation} 
In the special case that the phase $S$ does not depend on ${\bf x}$, for example for non-degenerate energy eigenstates, the total force is zero, so that the particle is free. In that case, particles will move to infinity unless their velocity is zero, which is the case for a trajectory of the \dbb\ theory. 

We can define the energy of a particle as 
\begin{equation}
{\widetilde E} = \frac{m |{\bf {\dot X}}|^2}{2} + V + Q \,,
\label{13.1}
\end{equation} 
where we have used the tilde to distinguish it from $E$, the energy of the wave function. We have that $d {\widetilde E}/dt = \pa Q / \pa t$, so that in general  $d {\widetilde E}/dt \neq 0$. However, in the case of an energy eigenstate, for which $Q$ does not depend on $t$, we have that $d {\widetilde E}/dt = 0$. For a trajectory given by the \dbb\ theory (i.e., corresponding to an initial velocity given by \eqref{5}), we have moreover that ${\widetilde E} = E$. (But ${\widetilde E} = E$ does not guarantee that the trajectory is given by the \dbb\ theory.) 

From \eqref{12}, it follows that the total potential $V + Q \leqslant E$. Hence, no matter how much the classical potential $V$ would confine the particles, the total potential is bounded from above. As such, it would seem that if the energy ${\widetilde E}$ of the particle were large enough, it would escape to infinity. Indeed, in the case of one spatial dimension, we have that when ${\widetilde E} > E(\geqslant V+Q)$, then the particle will escape to infinity, because there are no turning points. Actually, for a bound state in one dimension, we even have that $\pa_x S = 0$, so that $E=V+Q$ and the particle is free. 

Let us now turn to explicit examples.{\footnote{For the \dbb\ treatment of all the systems considered here, see \cite{holland93b}.}} We start with one spatial dimension. First consider a free particle. For a Gaussian wave function centered around the origin (we assume $\hbar=2m=\sigma=1$, where $\sigma$ is the width of the packet)
\be
\psi(x,t) = \left(\frac{1}{2\pi (1+\ii t)^2}  \right)^{1/4} \ee^{ -x^2/4(1 + \ii t)} \,,
\en
the QPD reads ${\ddot X} =  X/(1+t^2)^2$, so that the possible trajectories are 
\be
X(t) = \sqrt{1+t^2} (X_0 + V_0 \arctan t) \,,
\en
where $X_0$ and $V_0$ are respectively the initial position and velocity. The velocity constraint \eqref{5} corresponds to $V_0=0$. The standard deviation of the density $|\psi|^2$ is given by $\sigma(t) = \sqrt{1+t^2}$. Hence, if the initial speed $|V_0|$ of the particle is large enough, it will escape the bulk of the packet. For example, a particle with $0\leqslant X_0 < \sigma(0) = 1$ will have a position $X(t) > \sigma(t)$ for $t> \tan ((1 -X_0)/V_0)$ if $V_0 > 2(1 - X_0)/\pi$. In the case of an ensemble of particles with initial position and (independent) velocity distribution respectively given by $|\psi(x,0)|^2= \ee^{ -x^2/2}/\sqrt{2\pi}$ and $\ee^{ -(v-v_o)^2/2{\tilde \sigma}^2}/\sqrt{2\pi{\tilde \sigma}^2}$ (i.e., the initial velocity distribution is Gaussian with mean $v_0$ and standard deviation ${\tilde \sigma}$), the position distribution $\rho_{v_0}(x,t)$ at later times $t$ is Gaussian with mean $v_0 \sqrt{1+t^2}\arctan t$ and standard deviation $\sqrt{(1+t^2)(1 + {\tilde \sigma}^2 \arctan^2 t )}$. Hence, if $v_0 \not= 0$ the center of $\rho_{v_0}(x,t)$ moves away from that of the distribution $|\psi(x,t)|^2 = \rho_{v_0=0}(x,t)$, up to a distance $v_0\pi/2$ (after rescaling by $1/\sigma(t)$). $\rho_{v_0}$ also spreads more than $|\psi|^2$, by up to a factor $\sqrt{1+{\tilde \sigma}^2 \pi^2/4 }$.

In the case of a potential step of height $V$ for $x \geqslant 0$ (we assume $\hbar = 2m = 1$), the energy eigenstates $\psi(x,t) = \phi(x) \exp(-\ii Et)$ for $E < V$ are given by 
\begin{equation}
\phi (x) = \left\{ \begin{array}{ll} \cos(kx - \alpha/2) & x < 0\\ \cos(\alpha/2) \ee^{- \kappa x}   & x \geqslant 0 \end{array} \right. \,,
\label{14}
\end{equation} 
where $k= {\sqrt E}$, $\kappa = {\sqrt{V - E}}$ and $\alpha = 2\tan^{-1}(-\kappa/k)$. Since the phase of the wave function does not depend on the spatial coordinate, we have (as mentioned above) that the particle is free; it is not confined by the potential. (Actually, in this case, also the \dbb\ motion is unphysical, since particles will just stand still. A more realistic treatment should consider an approximately localized packet that moves towards the potential step.)

\begin{figure}
\centering
\includegraphics{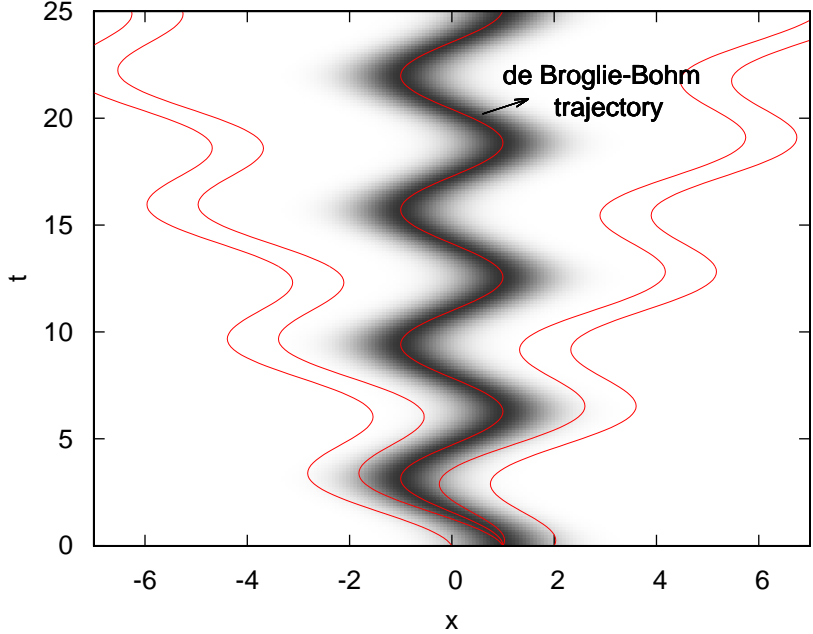}
\caption{Tractories for the QPD in the case of the coherent state \eqref{15.01} (with $a=1$). The density $|\psi|^2$ is represented by the shaded area.
 One \dbb\ trajectory is plotted, which follows the center of the density $|\psi|^2$. It is a trajectory with initial velocity $V_0 = 0$. The trajectories respectively moving to the right and left have initial velocities $V_0=\pm 0.25$ and escape the wave packet.}
\label{fig1}
\end{figure}

Consider now the harmonic oscillator, with $V= x^2/2$ (we assume $\hbar = m = \omega = 1$). The energy eigenstates are bound states and hence, as mentioned before, the particle is free. Other states of interest are coherent states
\be
\psi(x,t) = \pi^{-1/4} \exp \left[ -\frac{1}{2} (x-a \cos t)^2 - \frac{\ii}{2} \left( t +  + 2xa\sin t - \frac{a^2}{2} \sin 2t \right) \right] \,.
\label{15.01}
\en
The corresponding density
\be
|\psi(x,t)|^2 = \pi^{-1/2} \exp \left[ - (x-a \cos t)^2\right] 
\label{15.02}
\en 
is a Gaussian function whose center oscillates between the points $x=\pm a$. The QPD reads ${\ddot X} =  -a \cos t$, so that the possible trajectories are 
\be
X(t) = X_0 +  V_0t + a(\cos t -1) \,.
\label{15.03}
\en
Since $\pa S(x,0) / \pa x = 0$, the constraint \eqref{5} implies that $V_0 = 0$, so that the particle in the \dbb\ theory performs a harmonic oscillation around the point $X_0- a$. However, if $V_0 \neq 0$ then the particle will escape to infinity, oscillating around $X_0 + V_0t - a$. Some trajectories are plotted in figure \ref{fig1}.

\begin{figure}
\centering
\includegraphics{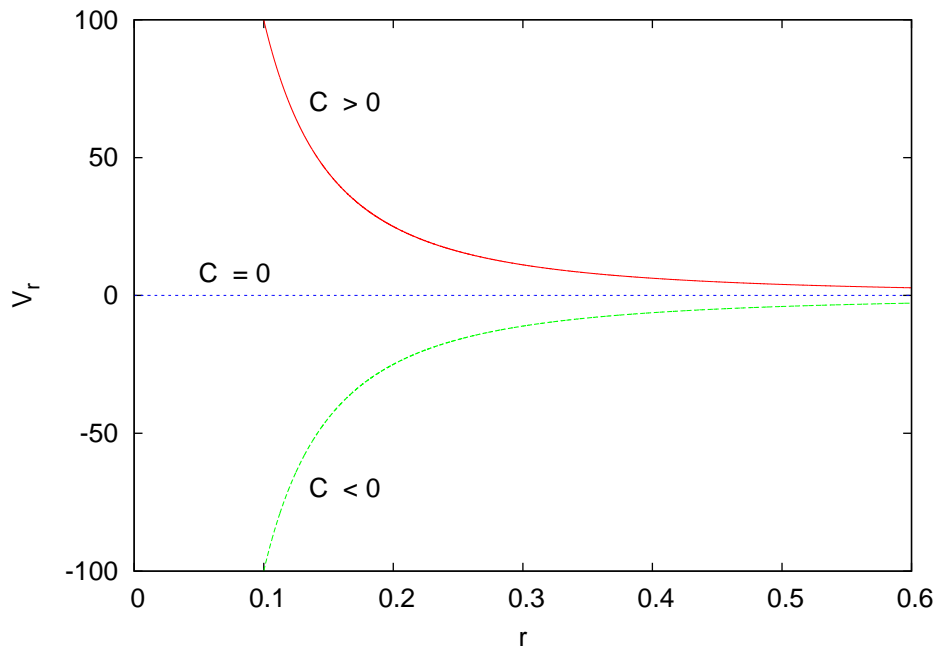}
\caption{The shape of the potential $V_r=C/ r^2$ for $C=\pm1,0$.}
\label{fig3}
\end{figure}

We now turn to three spatial dimensions and a central potential $V=V(r)$. Energy eigenstates with energy $E$, orbital angular momentum number $l$ and the magnetic quantum number $m$ (here we denote the mass by $m_0$) are of the form
\begin{equation}
\psi_{Elm}(r,\theta,\phi,t) = R_{El}(r) Y_{lm}(\theta,\phi)  \ee^{- \ii  Et/\hbar}  \,,
\label{16}
\end{equation} 
where $(r,\theta,\phi)$ are spherical coordinates determined by
\begin{equation}
x=r \sin \theta \cos \phi, \quad y=r \sin \theta \sin \phi, \quad z = r \cos \phi \,.
\label{17}
\end{equation} 
The functions $Y_{lm}$ are the spherical harmonics and $R_{El}$ are real functions that are solutions to the radial Schr\"odinger equation. Due to the spherical symmetry there is a $(2l+1)$-fold degeneracy: for given $E$ and $l$ all values of $m$ obeying $-l \leqslant m \leqslant l$ are possible. For bound states there is usually no further degeneracy so that general energy eigenstates are of the form
\be
\psi(r,\theta,\phi,t) = \sum^{l}_{m=-l} c_m \psi_{Elm}(r,\theta,\phi,t)  = R_{El}(r)  \sum^{l}_{m=-l} c_m Y_{lm}(\theta,\phi) \ee^{- \ii  Et/\hbar} \,,
\label{17.01}
\en 
where the $c_m$ are arbitrary complex coefficients. In the case of the harmonic potential and the Coulomb potential additional symmetries imply further degeneracy. In the following, we restrict our attention to states of the form \eqref{17.01}. Using \eqref{12} and the fact that the phase of $\psi$ does not depend on $r$, we find that for such a state the total potential is of the form 
\begin{equation}
V+Q = E - \frac{1}{2m_0} {\boldsymbol {\nabla}} S \cdot {\boldsymbol {\nabla}} S = E + \frac{1}{r^2} f(\theta,\phi) \,,
\label{18}
\end{equation} 
with $f=-r^2{\boldsymbol {\nabla}} S \cdot {\boldsymbol {\nabla}} S/2m_0$ a function of the angular variables only. So, dropping the constant $E$, the effective potential is of the form $f(\theta,\phi)/r^2$. This potential was considered in detail in \cite{wu94}. The corresponding Lagrangian is
\begin{equation}
L = \frac{m_0}{2}({\dot r}^2 + r^2 {\dot \theta}^2 + r^2 \sin^2 \theta {\dot \phi}^2) - \frac{f}{r^2} \,,
\label{19}
\end{equation}
with corresponding equations of motion
\be
m_0 {\ddot r} = m_0 r  ({\dot \theta}^2 + \sin^2 \theta {\dot \phi}^2) + 2 \frac{f}{r^3} \,,
\en
\be
\frac{d}{dt}(m_0  r^2 {\dot \theta} ) = m_0 r^2 \sin \theta \cos \theta {\dot \phi}^2 - \frac{1}{r^2} \frac{\pa f}{\pa \theta} \,,
\en
\be 
\frac{d}{dt}(m_0  r^2\sin^2 \theta{\dot \phi}) =  - \frac{1}{r^2} \frac{\pa f}{\pa \phi} \,.
\label{20}
\en
The energy ${\widetilde E}$ is a constant of the motion and reads (dropping again the constant $E$):
\begin{equation}
{\widetilde E} = \frac{m_0}{2}({\dot r}^2 + r^2 {\dot \theta}^2 + r^2 \sin^2 \theta {\dot \phi}^2) + \frac{f}{r^2}\,.
\label{22}
\end{equation}
In the case of a trajectory given by the \dbb\ theory we have that ${\widetilde E} = 0$. (We then also have that ${\dot r} = 0$, so that the particle moves on a sphere.) Another constant of the motion is 
\begin{equation}
C = \frac{1}{2m_0}|{\bf L}|^2 +  f  \,,
\label{21}
\end{equation}
where $|{\bf L}|^2 = m^2_0r^4({\dot \theta}^2 + \sin^2 \theta {\dot \phi}^2)$ with ${\bf L}$ the angular momentum. We can use it to write the energy as
\begin{equation}
{\widetilde E} = \frac{m_0}{2}{\dot r}^2 + \frac{C}{ r^2}.
\label{23}
\end{equation}
This shows that the coordinate $r$ moves under an effective potential $V_r = C/ r^2$, which is plotted in figure \ref{fig3}.

\begin{figure}
\centering
\includegraphics{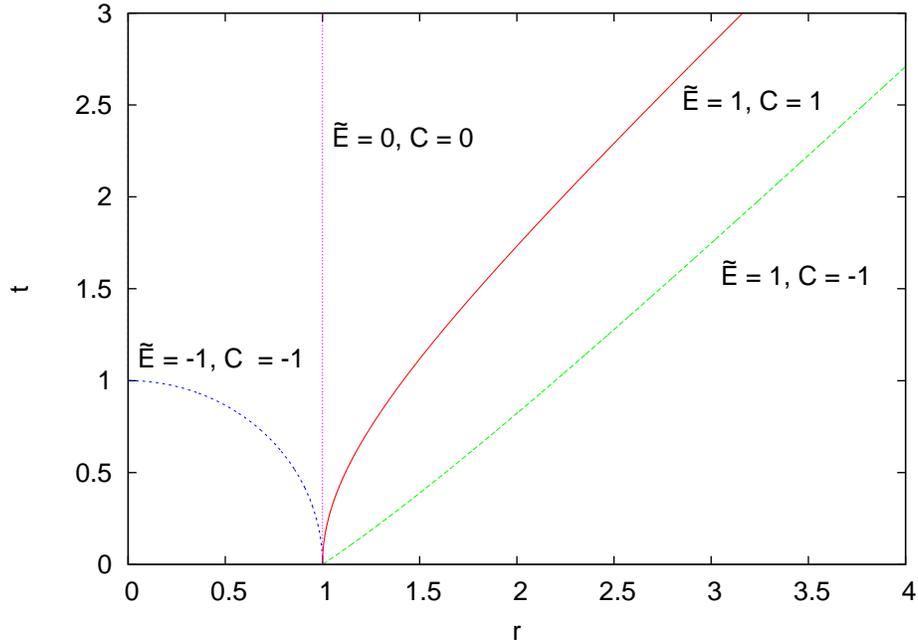}
\caption{Examples of radial trajectories with the same initial radial coordinate $r(0)=1$ (and $m_0=2$). If ${\widetilde E} > 0$ then the particle escapes to infinity.}
\label{fig4}
\end{figure}

We can now qualitatively analyze the QPD. In the case $C=0$, we have that ${\widetilde E} = m_0{\dot r}^2 / 2$. Hence, for strictly positive energy the particle flies off to infinity with a constant radial speed. For zero energy (as in the case of a \dbb-trajectory), the particle's trajectory is confined to the surface of a sphere. If $C>0$, we have ${\widetilde E} > 0$. The particle motion is then unbounded, with $r \geqslant (C/ {\widetilde E})^{1/2}$, and the particle eventually flies off to infinity. If $C< 0$, the particle motion is unbounded for ${\widetilde E} \geqslant 0$. For ${\widetilde E} < 0$ the particle motion is bounded, with $r \leqslant (C/ {\widetilde E})^{1/2}$. Examples of radial trajectories are presented in figure \ref{fig4}.

\begin{figure}
\centering
\includegraphics[scale=0.9]{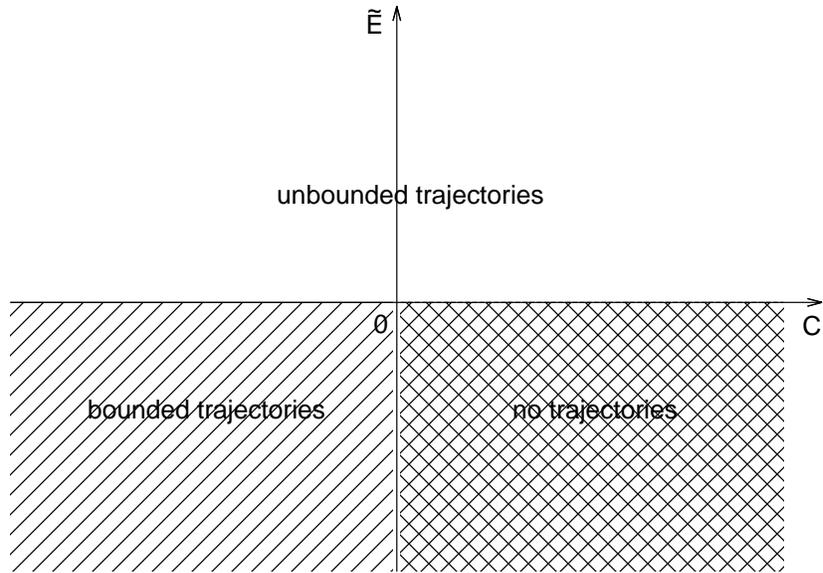}
\caption{Types of trajectories in terms of the constants of motion ${\widetilde E}$ and $C$. For points on the coordinate axes we have the following situation. For the origin, the motion is bounded. For $C>0$ (and ${\widetilde E} =0$) or ${\widetilde E}<0 $ (and $C=0$) we have no trajectories. For $C<0$ (and ${\widetilde E} =0$) the motion is unbounded.}
\label{fig2}
\end{figure}

The situation is summarized in figure \ref{fig2}. If the energy of the particle is greater than the energy in the \dbb\ theory, i.e., ${\widetilde E} > 0$, then its motion is unbounded and it flies off to infinity, even for initial conditions arbitrarily close to those of a trajectory of the \dbb\ theory. If ${\widetilde E} < 0$, then the particle motion will be bounded with radial motion oscillating between $0$ and the turning point $r=(C/ {\widetilde E})^{1/2}$, which may still be far from the center of the packet.

In conclusion, we have seen that according to the QPD particles often escape from the bulk of the wave packet, even for bound states. In the case of the hydrogen atom (which is described by the Coulomb potential), this means that the electron will not be bound to its nucleus. More generally we expect that molecules and atoms will not be stable but will tend to disintegrate. Of course we have only considered energy eigenstates here, but there seems to be no hope that the total potential will bind the particles in the case of a superposition (as was illustrated in \cite{colin13b}).{\footnote{As such, there is no point in even starting an analysis of positions measurements in QPD, in the hope that such measurements would yield the correct results, since measurement devices or pointers would not exist in the first place.}} Neither is there any hope that a relativistic or quantum field theoretical treatment will help. And, of course, even if this were not true, there would still seem to be no reason whatsoever to expect the predictions of QPD to be governed by Born probabilities and hence no reason whatsoever that they should have anything to do with those of quantum mechanics.

This work might also be of relevance for particular techniques for simulating the wave function evolution using \dbb\ trajectories. For example, one technique is roughly as follows \cite{lopreore99,wyatt05,deckert07,goldstein11a}. The Newtonian equation \eqref{3} is considered with $|\psi|$ replaced by ${\sqrt \rho}$, where $\rho$ an actual density of a large but finite number of configurations. Given an initial wave function $\psi_0$, the equation is numerically integrated starting with the initial distribution $\rho_0$ given by $|\psi_0|^2$ (up to some accuracy) and with the initial velocity of each configuration satisfying the constraint \eqref{5}. From the trajectories one can then obtain a time-dependent wave function $\psi$. But since $\rho$ equals $|\psi|^2$ only up to some accuracy, the simulation will entail small deviations of the velocities from $\nabla S/m$. As is illustrated by our results, this might cause trajectories to deviate significantly from trajectories in the \dbb\ theory and hence this may potentially lead to an inaccurate simulation of the wave function. (It may also be that the difference between the force determined by $\rho$ and the quantum force determined by $|\psi|^2$ induces corrections to the velocity which brings them closer to $\nabla S/m$.)

Recently, this particular way of numerically simulating the wave function formed the basis of a new approach to quantum mechanics \cite{hall14,sebens14}. In this approach, there is a large but finite number of configurations, each representing a different world, which evolve according to a dynamics similar to that used in the wave function simulations. There is no wave function on the fundamental level. But given appropriate initial conditions on the velocities, the configurations may (approximately) determine a wave function. However, it is not clear whether such wave functions obey some Schr\"odinger dynamics. One potential source of trouble may be the instabilities caused by deviations of the velocities from $\nabla S/m$.

{\bf Acknowledgments.} It is a pleasure to thank Roderich Tumulka for discussions. S.G.~and W.S.~are supported in part by the John Templeton Foundation. The opinions expressed in this publication are those of the authors and do not necessarily reflect the views of the John Templeton Foundation.

\end{document}